\documentclass[aps,prl,twocolumn,superscriptaddress]{revtex4-1}
\usepackage{textcomp}
\usepackage{graphicx}
\usepackage{longtable}

\begin{document}
\title{Structural, Magnetic, and Superconducting Properties of Ba$_{1-x}$Na$_{x}$Fe$_{2}$As$_{2}$}
\author{S. Avci}
\affiliation{Materials Science Division, Argonne National Laboratory, Argonne, IL 60439-4845}
\author{J. M. Allred}
\thanks{Corresponding Author}
\affiliation{Materials Science Division, Argonne National Laboratory, Argonne, IL 60439-4845}
\author{O. Chmaissem}
\affiliation{Materials Science Division, Argonne National Laboratory, Argonne, IL 60439-4845}
\affiliation{Physics Department, Northern Illinois University, DeKalb, IL 60115}
\author{ D. Y. Chung}
\affiliation{Materials Science Division, Argonne National Laboratory, Argonne, IL 60439-4845}
\author{S. Rosenkranz}
\affiliation{Materials Science Division, Argonne National Laboratory, Argonne, IL 60439-4845}
\author{J. A. Schlueter}
\affiliation{Materials Science Division, Argonne National Laboratory, Argonne, IL 60439-4845}
\author{ H. Claus}
\affiliation{Materials Science Division, Argonne National Laboratory, Argonne, IL 60439-4845}
\author{A. Daoud-Aladine}
\affiliation{ISIS Pulsed Neutron and Muon Source, Rutherford Appleton Laboratory, Chilton, Didcot OX11 0QX, United Kingdom}
\author{D. D. Khalyavin}
\affiliation{ISIS Pulsed Neutron and Muon Source, Rutherford Appleton Laboratory, Chilton, Didcot OX11 0QX, United Kingdom}
\author{P. Manuel}
\affiliation{ISIS Pulsed Neutron and Muon Source, Rutherford Appleton Laboratory, Chilton, Didcot OX11 0QX, United Kingdom}
\author{A. Llobet}
\affiliation{Lujan Neutron Scattering Center, Los Alamos National Laboratory, Los Alamos, NM 87545, USA}
\author{M. R. Suchomel}
\affiliation{Advanced Photon Source, Argonne National Laboratory, Argonne Illinois 60439, USA}
\author{M. G. Kanatzidis}
\affiliation{Materials Science Division, Argonne National Laboratory, Argonne, IL 60439-4845}
\affiliation{Department of Chemistry, Northwestern University, Evanston, IL 60208-3113, USA}
\author{R. Osborn}
\affiliation{Materials Science Division, Argonne National Laboratory, Argonne, IL 60439-4845}
\date{\today}

\begin{abstract}
We report the results of a systematic investigation of the phase diagram of the iron-based superconductor system, Ba$_{1-x}$Na$_{x}$Fe$_{2}$As$_{2}$, from \emph{x} = 0.1 to \emph{x} = 1.0 using high resolution neutron and x-ray diffraction and magnetization measurements. We find that the coincident structural and magnetic phase transition to an orthorhombic (O) structure with space group $Fmmm$ and a striped antiferromagnet (AF) with space group $F_Cmm'm'$  in Ba$_{1-x}$Na$_{x}$Fe$_{2}$As$_{2}$  is of first order. A complete suppression of the magnetic phase is observed by \emph{x} = 0.3, and bulk superconductivity occurs at a critical concentration near 0.15.  We compare the new findings to the previously reported results of the hole-doped Ba$_{1-x}$K$_{x}$Fe$_{2}$As$_{2}$ solid solution in order to resolve the differing effects of band filling and A-site cation size on the properties of the magnetic and superconducting ground states. The substantial size difference between Na and K causes various changes in the lattice trends, yet the overarching property phase diagram from the Ba$_{1-x}$K$_{x}$Fe$_{2}$As$_{2}$ phase diagram carries over to the Ba$_{1-x}$Na$_{x}$Fe$_{2}$As$_{2}$ solid solution. We note that the composition dependence of the \emph{c} axis turns over from positive to negative around \emph{x} = 0.35, unlike the K-substituted materials.  We show that this can be understood by invoking steric effects; primarily the Fe$_{2}$As$_{2}$ layer shape is dictated mostly by the electronic filling, which secondarily induces an interlayer spacing adjusted to compensate for the given cation volume. This exemplifies the primacy of even subtle features in the Fe$_{2}$As$_{2}$ in controlling both the structure and properties in the uncollapsed 122 phases.
\end{abstract}
\maketitle
\section{Introduction}
Superconductors derived from the ‘parent’ compound BaFe$_{2}$As$_{2}$ are amongst the most heavily studied of the iron arsenides over the last four years \cite{1,2}. BaFe$_{2}$As$_{2}$ crystallizes in the so-called ‘122’ tetragonal structure at room temperature but orders antiferromagnetically below 139 K with a simultaneous structural transition to orthorhombic symmetry \cite{3,4,5}. Both the orthorhombic and tetragonal structures contain Fe$_2$As$_2$ layers, in which the iron atoms are tetrahedrally coordinated with arsenic atoms.  As with the other iron-based superconductors, the suppression of spin density wave (SDW) order and the emergence of superconductivity can be controlled by electron- or hole-doping through aliovalent substitution on the barium or iron sites\cite{5,6,7,8}, by isovalent substitution on the arsenic sites\cite{9,10}, or by the application of external or chemical pressure \cite{11,12}.

In the electron-doped systems, such as Ba(Fe$_{1-x}$Co$_{x}$)$_{2}$As$_{2}$ and Ba(Fe$_{1-x}$Ni$_{x}$)$_{2}$As$_{2}$, the maximum \emph{T}$_{c}$ is ~25 K and superconductivity vanishes at only 0.12 electrons/Fe atom \cite{7}.  It is well established that the magnetic and structural transitions are separated and that superconductivity and magnetism coexist microscopically as shown by local probes such as muon spin rotation (μSR) \cite{13} or nuclear magnetic resonance (NMR) measurements \cite{14,15}. On the other hand, superconductivity in the hole doped Ba$_{1-x}$K$_{x}$Fe$_{2}$As$_{2}$ system extends up to 0.5 holes/Fe atom, i.e., to KFe$_{2}$As$_{2}$, with a maximum \emph{T}$_{c}$ of 38 K, significantly higher than the electron-doped materials \cite{5,8}.  An important advantage is that the doping is intrinsically cleaner; since the potassium substitution on the barium sites leaves intact the superconducting Fe$_{2}$As$_{2}$ layers.  The microscopic coexistence of magnetism and superconductivity in Ba$_{1-x}$K$_{x}$Fe$_{2}$As$_{2}$ was recently established by neutron diffraction \cite{4,5} and μSR measurement \cite{16}. In Ref. \cite{4}, we also reported a strong magneto-elastic coupling in Ba$_{1-x}$K$_{x}$Fe$_{2}$As$_{2}$ as shown by the coincident first-order magnetic and structural transitions and the identical temperature dependence of the magnetic and orthorhombic order parameters. We discussed the theoretical implications of this coupling and the origin of the electron-hole asymmetry in the phase diagram. Therefore, it is essential to determine whether this is a common property for all hole-doped 122 systems or that the K substituted system may constitute a special case perhaps due to the nearly similar ionic sizes of eight coordinated barium (1.56 \r{A}) and potassium (1.65 \r{A}). In this paper, we report similar findings for the alternative hole-doped Ba$_{1-x}$Na$_{x}$Fe$_{2}$As$_{2}$ system in which the sodium ions (1.32 \r{A}) are significantly smaller than barium ions \cite{17}.

In recent work, Cortes-Gil \emph{et al} reported the magnetic and superconducting properties of a few samples of Ba$_{1-x}$Na$_{x}$Fe$_{2}$As$_{2}$ \cite{18}.  The authors observed magnetic ordering and an orthorhombic structural distortion for \emph{x} ≤ 0.35 and bulk superconductivity for samples within the range 0.4 < \emph{x} <0.6, with a maximum \emph{T}$_{c}$ of 34 K occurring at \emph{x} = 0.4.  Additionally, they determined an upper limit for the Na substitution at \emph{x}=0.6 and did not observe any cation ordering between Ba$^{2+}$ and Na$^{1+}$ ions. However, the onset of superconductivity is not well-defined. There is evidence of a reduced superconducting phase fraction of ~30\% at \emph{x} = 0.3, but it is not clear if this is intrinsic or due to chemical inhomogeneity. The researchers reported broad superconducting transitions and noisy lattice parameters, which probably resulted from the difficulty in controlling the synthesis chemically because of the large size mismatch between Ba and Na and the volatility of the latter at high temperatures.  Sodium has also been used as a dopant in the other divalent iron arsenides (Sr, Ca, Eu) to successfully induce superconductivity. \cite{19,20,21}

In this paper, we report a method for synethesizing high quality, stoichiometric Ba$_{1-x}$Na$_{x}$Fe$_{2}$As$_{2}$ and  a detailed, revised phase diagram  determined by neutron and x-ray powder diffraction and magnetization measurements. By maintaining Na stoichiometry we demonstrate that superconductivity is established at around 1 K in the \emph{x} = 0.15 sample suggesting that the superconductivity dome must begin around this compositional value.   There is no upper substitution limit since all materials up to nominally NaFe$_{2}$As$_{2}$ (\emph{T}$_{c}$ = 12 K) were found superconducting \cite{22}. To a great extent, we find that Ba$_{1-x}$Na$_{x}$Fe$_{2}$As$_{2}$  and  Ba$_{1-x}$K$_{x}$Fe$_{2}$As$_{2}$  are quite similar including the fact that both systems exhibit coincident first-order magnetic and orthorhombic transitions with a strong magneto-elastic coupling. A small region of phase coexistence of magnetism and superconductivity is also observed.

\section{Experimental}
The synthesis of homogeneous single phase Ba$_{1-x}$Na$_{x}$Fe$_{2}$As$_{2}$ polycrystalline powders was performed using conditions that are similar to Ba$_{1-x}$K$_{x}$Fe$_{2}$As$_{2}$. Details of the synthesis can be found in Ref. \cite{5}.  Briefly, handling of all starting materials was carried out in a N$_{2}$-filled glove box.  Mixtures of Ba, Na, and FeAs were loaded in alumina tubes, sealed in niobium tubes under argon, and sealed again in quartz tubes under vacuum. The mixtures were heated at 850\textdegree~C for one day and ground to homogeneous powders and then reheated for 3 days at 850\textdegree~C. This last step was performed twice. Before the last anneal a slight amount of NaAs was added. The structure and quality of the final black powders were confirmed by x-ray powder diffraction and magnetization measurements.  For select samples, neutron powder diffraction experiments were performed on the High Resolution Powder (HRPD) (\emph{x} = 0) and WISH Diffractometers (\emph{x} = 0.15) at ISIS (Rutherford Appleton Laboratory) and on the High Intensity Powder Diffractometer (HIPD) (\emph{x} = 0.1, 0.2, 0.3) at LANSCE (Los Alamos National Laboratory). Synchrotron x-ray diffraction measurements were performed as a function of temperature for \emph{x} = 0, 0.1, 0.15, 0.2, 0.3, 0.4, 0.5, 0.6, 0.7, 0.8 on the high resolution powder diffractometer 11BM-B at the Advanced Photon Source of Argonne National Laboratory. The \emph{x} = 0.15 samples measured at 11BM-B and WISH are from two different batches. The nuclear and magnetic structures and internal parameters were determined by the Rietveld technique using GSAS \cite{23} and EXPGUI \cite{24}.
\section{Results and Discussion}
\subsection{Superconductivity and sample quality}
 \begin{figure}[b]
 \includegraphics[width=8.6cm]{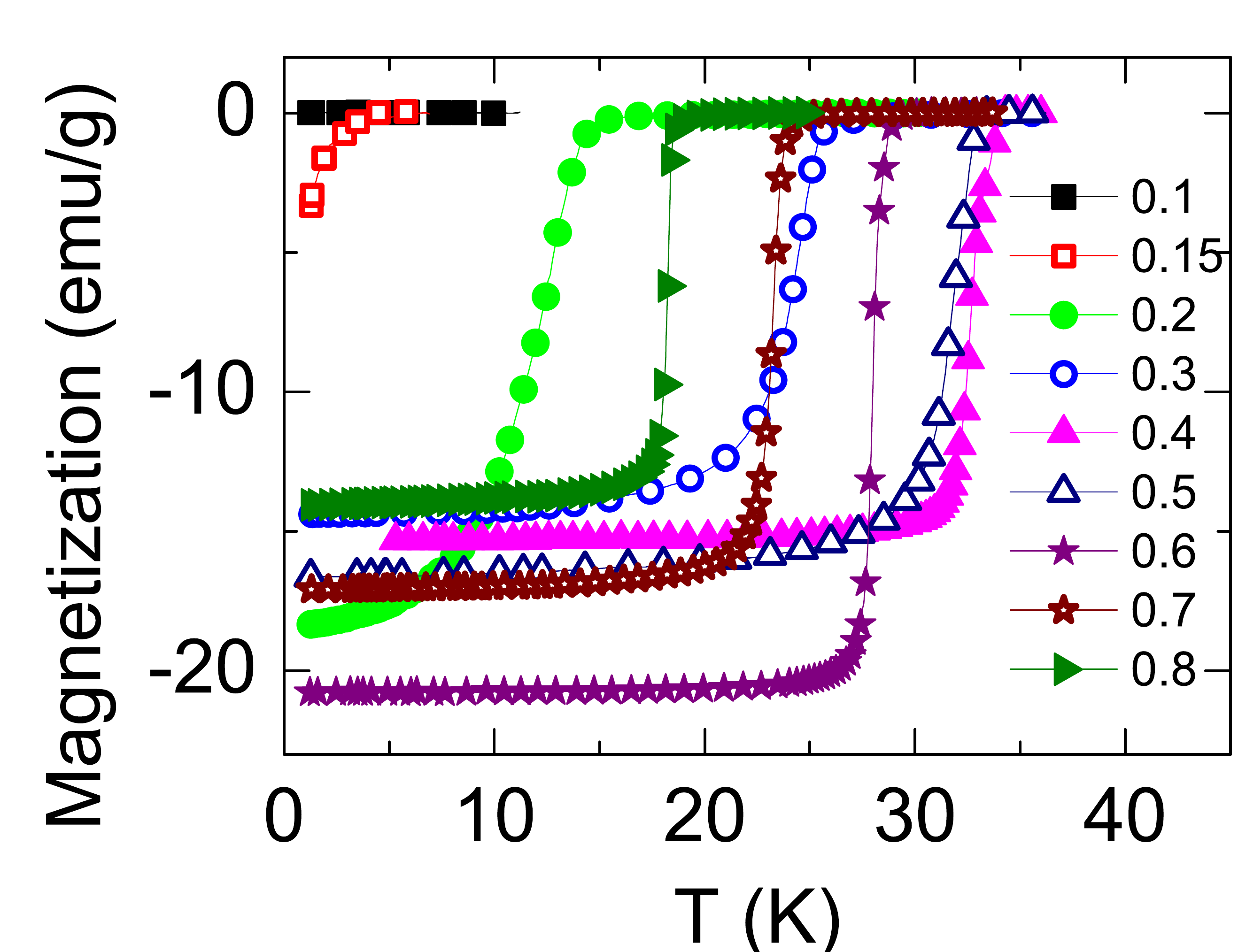}
 \caption{Magnetization measurements of Ba$_{1-x}$Na$_x$Fe$_2$As$_2$ system. All the samples were measured in a 0.1G applied magnetic field. Magnetization values are normalized to the mass of the samples.\label{Figure1}}
 \end{figure}
As is common with light alkali metal containing compounds, repeated anneals of these sodium doped materials cause degradation of sample quality, observed in both the x-ray diffraction peak shapes and magnetic measurements of the Meissner effect. Annealing steps were kept as short as possible, though in order to accommodate just long enough anneal times to get chemically homogeneous powders, the inclusion of excess Na in the form of NaAs was used. Fig. 1 shows zero\textendash field\textendash cooled magnetic measurements and the evolution of superconductivity as a function of Na content. All the samples were measured in a 0.1 Oe applied magnetic field.  The magnetization curves are normalized to the mass of the samples. \emph{T}$_{c}$ is determined from the onset of the divergences of the first derivative of the magnetic susceptibility (1.4, 15.2, 25.8, 34.2, 33.2, 28.7, 24, and 18.6 K for \emph{x} = 0.15, 0.2, 0.3, 0.4, 0.5, 0.6, 0.7, and 0.8, respectively). As seen in the figure, trace superconductivity below 3 K is observed for \emph{x} = 0.15 that only starts to diverge (\emph{T}$_{c}$ onset) below 1.4 K (just above our instrumental limit of 1.3 K). Extrapolating a similar curve width as other samples shows that this sample reaches $\approx 50\%$ volume fraction around 1 K, meaning that the 15\% composition exists just above the superconducting dome phase line. We lack finely spaced compositions, but we can note that the critical temperature initially increases rapidly along with the Na concentration to a \emph{T}$_{c}$ onset of 15 K for \emph{x} = 0.20, and then above this composition the compositional dependence becomes weaker. This shape for the underdoped superconducting dome phase line is somewhat different from what we previously observed in the Ba$_{1-x}$K$_{x}$Fe$_{2}$As$_{2}$ system: an onset of \emph{x} = 0.13 and linear compositional dependence of \emph{T}$_{c}$.\cite{5}

Upon further increasing \emph{x} in Ba$_{1-x}$Na$_{x}$Fe$_{2}$As$_{2}$, \emph{T}$_{c}$ rises to a maximum of 34 K for \emph{x} = 0.4 and then gradually falls back to 12 K for the end composition\emph{x} = 1, which was determined previously \cite{22}. Our values further refine the shape and position of the superconducting dome; note that we find that \emph{T}$_{c}$ onsets at significantly lower compositions than previously reported \cite{18}.  It is likely that the actual composition in Ref \cite{18} was less than nominal, as discussed above. The $x \geq 0.8$ samples are superconducting (only \emph{x} = 0.8 is shown) but as noted in Ref. \cite{18} these samples are metastable in air and partly degrade.  For these samples no structural data will be reported, because synchrotron x-ray diffraction data showed significant amounts of impurities, and the chemical decomposition affected the crystallinity and stoichiometry enough to make Rietveld analysis unreliable.

\subsection{Structure}
 \begin{figure}
 \includegraphics[width=8.6cm]{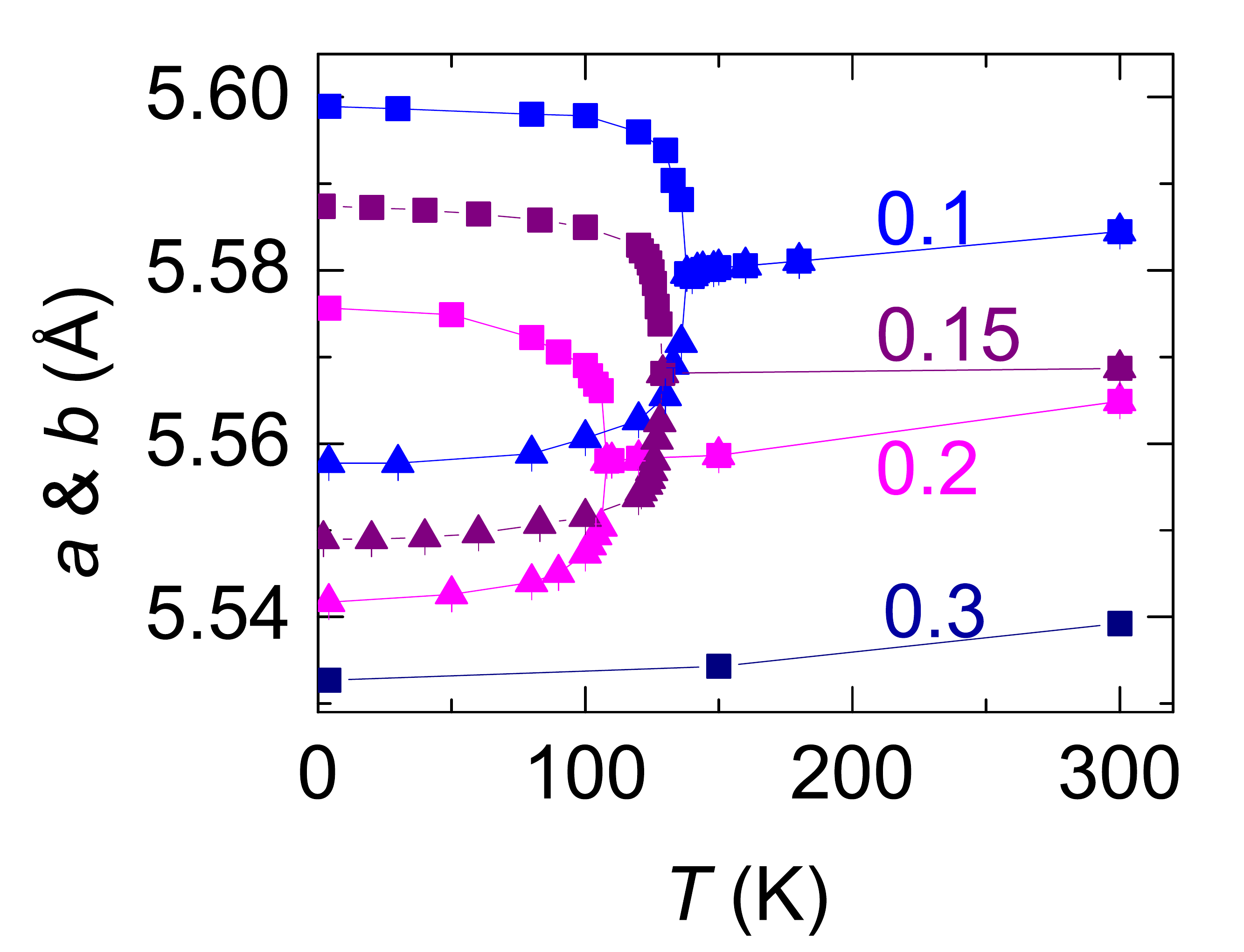}
\vspace{-15pt}
 \caption{Variation of lattice constants \emph{a} and \emph{b} as a function of temperature for \emph{x} = 0.1, 0.15, 0.2 and 
0.3 samples determined by neutron diffraction. The lattice constants in the tetragonal phase are multiplied by $\sqrt{2}$.\label{Figure2}}
 \end{figure}
The structure of both the tetragonal and orthorhombic phases of Ba$_{1-x}$Na$_x$Fe$_2$As$_2$ can be described as a stack of edge-sharing Fe$_{2}$As$_{2}$ layers separated by layers of \emph{A}-site (Ba, Na) ions as in the Ba$_{1-x}$K$_{x}$Fe$_{2}$As$_{2}$ system. In the undoped BaFe$_{2}$As$_{2}$ compound, a structural phase transition takes place at 139 K from the tetragonal ThCr$_2$Si$_2$-type structure of space group I4/mmm to an orthorhombic $\beta -$SrRh$_{2}$As$_{2}$-type structure of space group $Fmmm$ \cite{8}.  In Ba$_{1-x}$Na$_{x}$Fe$_{2}$As$_{2}$, the $x \leq 0.24$ samples experience similar tetragonal to orthorhombic structural transitions evidenced by the splitting of the basal \emph{a} and \emph{b} lattice parameters in the orthorhombic cell (Fig. 2). The transition temperature, $T_s$, decreases with increasing the Na content and for \emph{x} = 0.3, the transition is completely suppressed. We have seen evidence \cite{25} that at compositions just below complete suppression of magnetism (starting at \emph{x} = 0.24), a reentrant antiferromagnetic tetragonal (AF/T or “C4”) transition around 40 K, with a different magnetic structure than the antiferromagnetic orthorhombic (AF/O) phase. It should also be noted that the AF/T becomes stable well above the superconducting transition (40 K vs. 18 K).  Details of this new phase and its implications for the origin of the magneto-structural transitions are the subject of a separate report, but we have included it in this phase diagram for completeness. 

 \begin{table*}
 \caption{Lattice parameters of Ba$_{1-x}$Na$_x$Fe$_2$As$_2$ determined from refinements of synchrotron x-ray (11BM-B) and neutron (HIPD and WISH) powder diffraction data. For the orthorhombic samples volume is given for the tetragonal subcell. In these cases the actual volume is twice what is reported. \label{Table I}}
 \begin{ruledtabular}
 \begin{tabular}{c c c c c c}
\emph{x} & \emph{a} (\r{A}) & \emph{b} (\r{A}) & \emph{c} (\r{A}) & \emph{V} (\r{A}$^3$) & $z_{As}$ \\
\hline
\multicolumn{6}{c}{Neutrons, 4 K} \\
\hline
0.1 & 5.5989(3) & 5.5578(3) & 12.9913(3) & 202.129(15) & 0.35457(17) \\
0.15 & 5.5874(3) & 5.5490(3) & 12.9806(3) & 201.228(16) & 0.35489(6) \\
0.2 & 5.5757(3) & 5.5417(3) & 13.0184(6) & 201.125(18) & 0.35508(15) \\
0.3 & 3.9122(2) & - & 13.0393(6) & 199.572(15) & 0.35561(17) \\
\hline
\multicolumn{6}{c}{X-rays, 8 K}  \\
\hline
0.1 & 5.58558(7) & 5.54539(7) & 12.96689(13) & 200.820(4) & 0.35472(6) \\
0.15 & 5.57927(6) & 5.53967(6) & 12.97732(13) & 200.547(4) & 0.35457(7) \\
0.2 & 5.56117(10) & 5.53023(9) & 12.98378(18) & 199.656(6) & 0.35556(6) \\
0.3 & 3.90312(2) & - & 13.01261(10) & 198.238(2) & 0.35622(4) \\
0.4 & 3.88672(3) & - & 13.02818(10) & 196.812(3) & 0.35687(4) \\
0.5 & 3.86920(3) & - & 13.01707(11) & 194.875(3) & 0.35717(3) \\
0.6 & 3.85330(5) & - & 12.97904(18) & 192.712(4) & 0.35886(6) \\
0.7 & 3.83615(5) & - & 12.9217(3) & 190.155(5) & 0.35906(5) \\
 \end{tabular}
 \end{ruledtabular}
 \end{table*}

The low temperature lattice parameters as determined from Rietveld analysis of both the x-ray and neutron experiments are each reported in Table I. Fig. 3 shows the temperature dependence of the tetragonal (110) Bragg peak for \emph{x} = 0.1, 0.15, 0.2, and 0.3 in the vicinity of the structural transition. This reflection splits into the (022) and (202) orthorhombic peaks below $T_s$ for \emph{x} = 0.1, 0.15 and 0.2 consistent with the splitting of the \emph{a} and \emph{b} cell parameters.  The \emph{x} = 0.3 sample is tetragonal down to 4 K as shown in Fig. 3. For the \emph{x} = 0.2 sample, the splitting of the (110) peak is not resolved but a significant peak broadening is observed.  Close to the transition, the two peaks cannot be resolved but we can determine \emph{T}$_{s}$ from the temperature dependence of the full-width-at-half-maximum (FWHM) which becomes constant for the high temperature tetragonal phase. The transition temperatures were also confirmed by fitting the temperature–dependent orthorhombic order parameter $\delta = (a-b/a+b)$ to a power law $(T_s - T)^{\beta}/T_s$ for \emph{x} = 0, 0.1, 0.15 and 0.2 samples giving ~138, 137, 128, and 107 K with exponents $\beta$ = 0.13, 0.21, 0.20, and 0.19, respectively. These values are similar to those observed in the K-doped system. These values do not necessarily reflect the intrinsic critical exponents, which will be modified by slight chemical inhomogeneity.

 \begin{figure}
 \includegraphics[width=8.6cm]{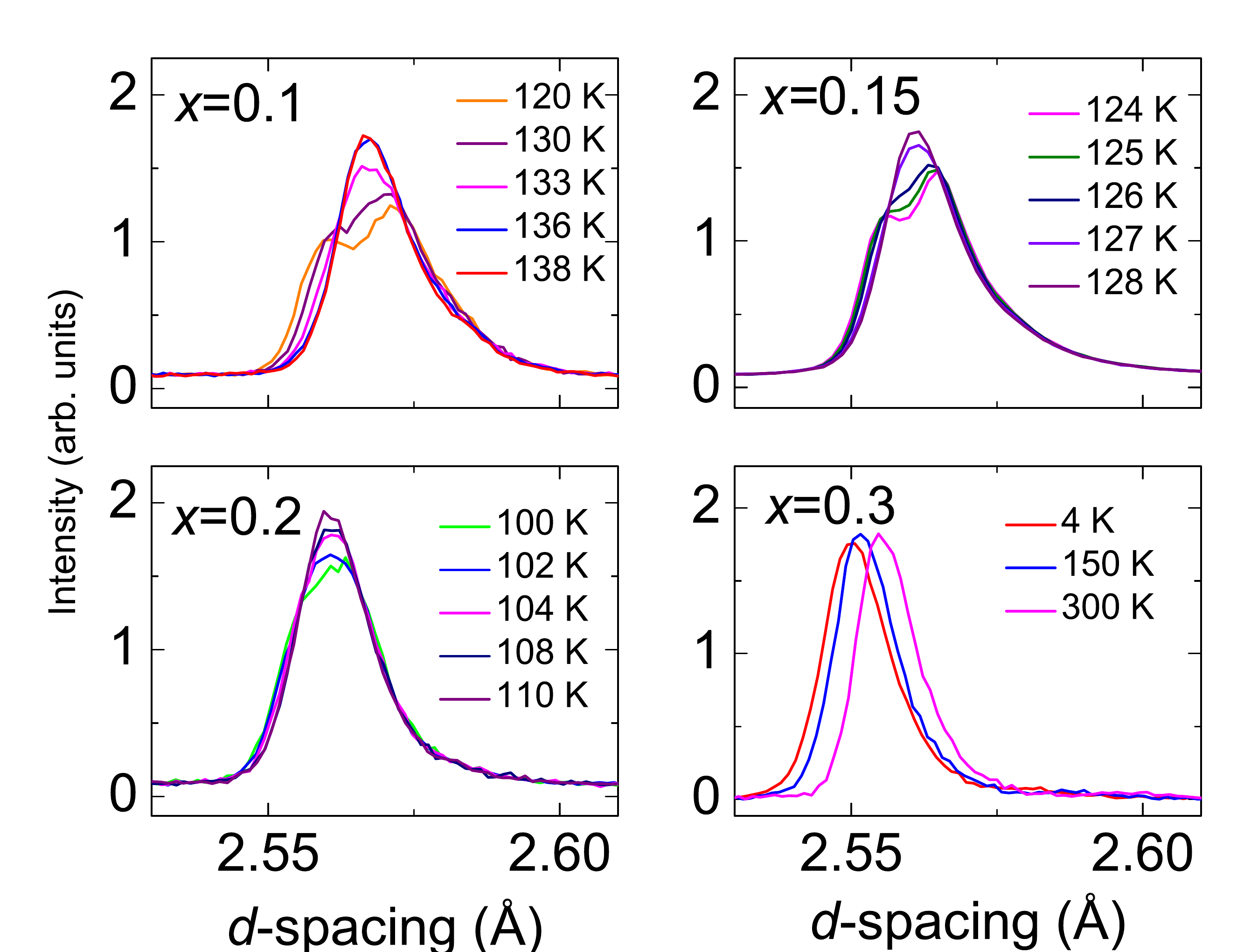}
 \caption{Temperature dependence of the (110) peak measured by neutron diffraction in the vicinity of structural transition temperature $T_s$ for \emph{x} = 0.1, 0.15, 0.2. The bold curve shows the peak at the estimated $T_s$. The \emph{x} = 0.3 sample remains tetragonal down to 4 K. \label{Figure3}}
 \end{figure}

 \begin{figure}
\vspace{10pt}
 \includegraphics[width=8.4cm]{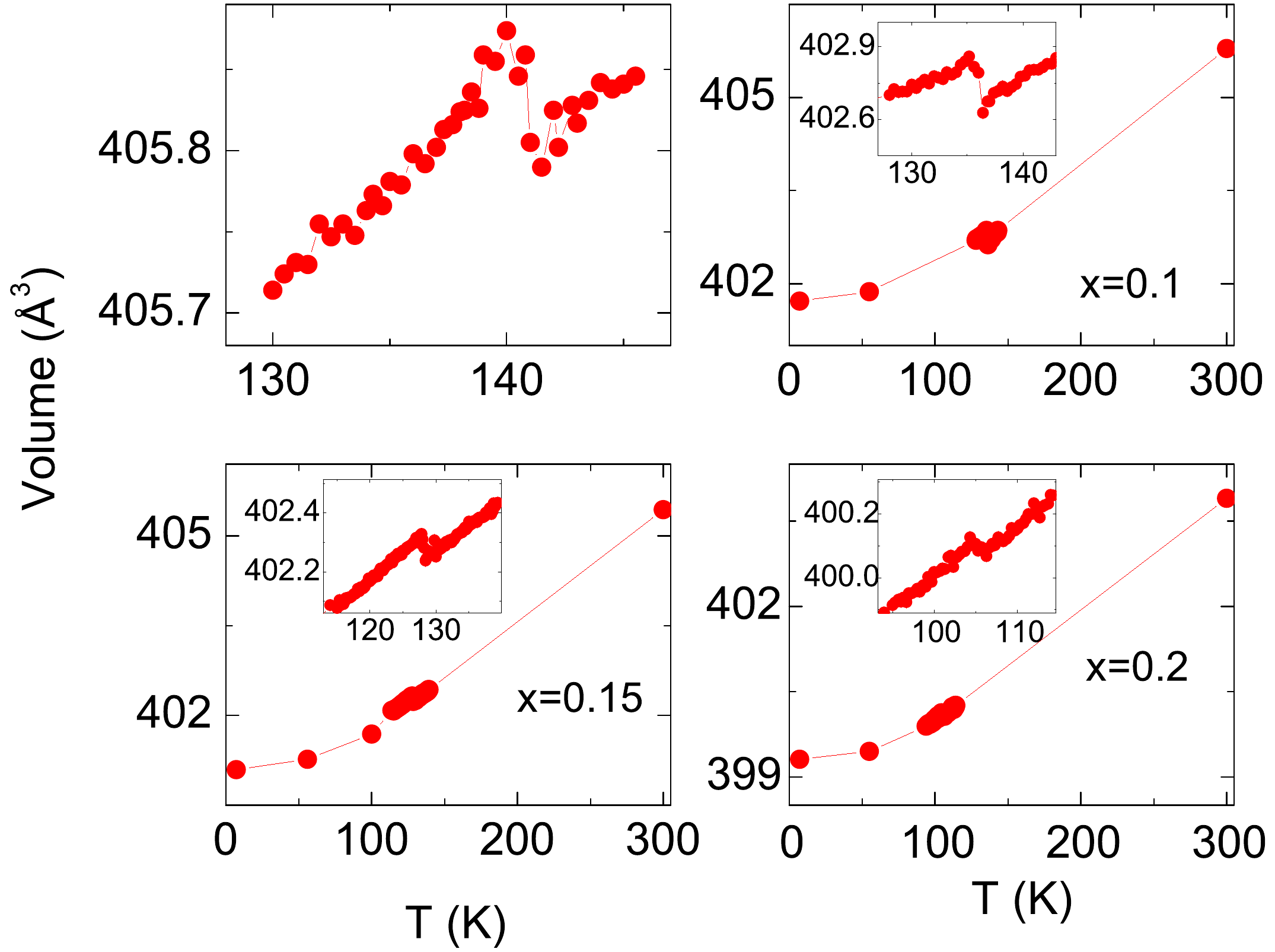}
\vspace{-1pt}
 \caption{Temperature dependence of the unit cell volume for \emph{x} = 0, 0.1, 0.15 and 0.2 determined by Rietveld refinements using synchrotron x-ray data. \label{Figure4}}
 \end{figure}

We previously showed small but sharp volume anomalies, a clear signature of a first order transition, in the vicinity of $T_s$ for the hole-doped Ba$_{1-x}$K$_x$Fe$_{2}$As$_{2}$ ($x \leq 0.24$).  Unfortunately, due to the limited neutron beam time, we were not able to collect enough data to show similar anomalies; however, x-ray data at 11BM-B showed clear volume jumps at the structural transition temperatures for \emph{x} = 0.1, 0.15 and 0.2 samples as shown in Fig. 4.  It must be emphasized that such anomalies were not observed in the electron doped Co, Rh, Ni and Pd substituted Ba-122 systems \cite{26}.  In fact, the structural and magnetic transitions in these latter systems are separated and the structural transitions are always second order \cite{27}. Our results of first-order like transitions in both the Na and K substituted systems, regardless of ionic size, may either be related to the intrinsic cleanliness of these systems or to a difference with hole-doping \cite{4,5}. In these systems, substituting Ba with K or Na leaves the Fe$_{2}$As$_{2}$ layers undisturbed.  The evolution of magnetism in these layers is not hindered by substitution-induced localized defects or distortions and the magnetoelastic coupling is therefore strong.

 \begin{figure}
 \includegraphics[width=8.6cm]{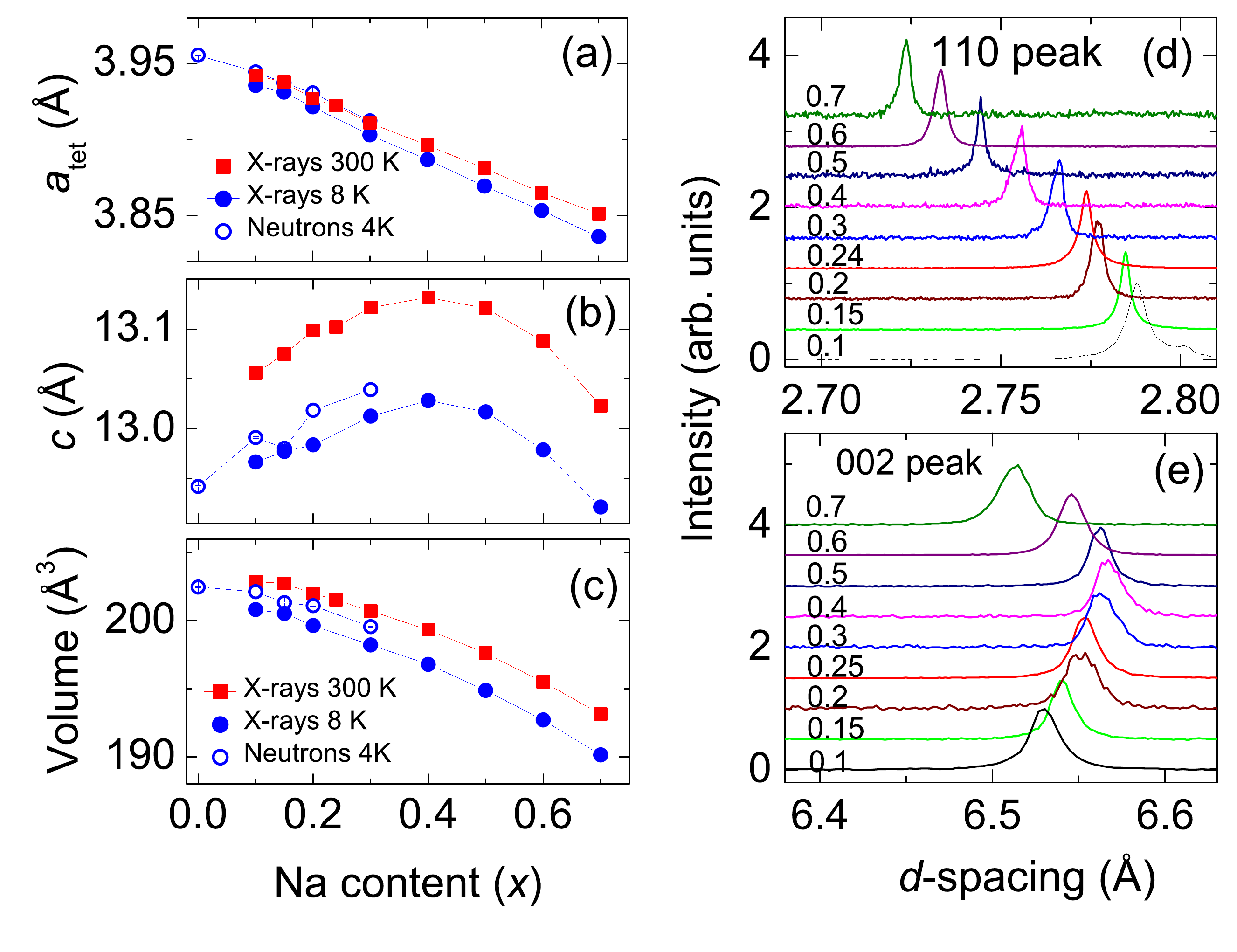}
\vspace{-20pt}
 \caption{ (a) Na content dependence of \emph{a} extracted from neutron (4 K) and x-ray (8 K) data. For orthorhombic samples, $a_{tet}  =  \sqrt{a^2+b^2 }/2$. (b) the \emph{c}-axis and (c) unit cell volume as a function of increasing Na content extracted from neutron (4 K, open blue circles) and x-rays (300 K, red squares and 8 K, solid blue circles) data. (d and e) 11BM-B data collected at room temperature showing the tetragonal (110) and (002) peaks for various Na concentrations. \label{Figure5}}
 \end{figure}

The refined lattice parameters are shown in Fig. 5; the obtained values vary smoothly which is indicative of good chemical control. Fig. 5(a) shows the evolution of the basal plane tetragonal a lattice parameter as a function of Na content obtained using neutrons (4 K) and x-rays (8 K and 300K), which differ slightly due to instrumental effects. In Fmmm the orthorhombic $a$ and $b$ lattice parameters can be transformed to give the I4/mmm tetragonal spacing by using the crystallographic relation $a_{tet}=\sqrt{a^2 + b^2}/2$. The \emph{a}-axis is weakly temperature dependent and decreases monotonically as expected by a naïve size comparison but the behavior of the \emph{c}-axis is remarkable in that it shows a bell shaped response to the increasing Na content peaking at $x\approx 0.4$. A clear evidence for this behavior is also seen in the raw data of the (002) peak shown in Fig. 5(e). This is rather different than other systems such as the K-doped, where the c-axis increases monotonically, and the Co-doped, where there are only small changes in lattice parameters. Generally, one would expect behavior that follows Vegard’s Law: constant change in volume and lattice parameters scaled by the two end points. In this case, the Na ion is significantly smaller than Ba (1.32 vs 1.65 \r{A}), indeed so much so that the substitution constitutes a significant strain. This is related to why another form, NaFeAs (the 111 phase), has an extra layer of alkali atoms inserted between the Fe$_{2}$As$_{2}$ layers.  The \emph{A} site void size is set in part by the Fe$_{2}$As$_{2}$ layer spacing and is relaxed in the 111 structure by introducing twice as many A sites, which are staggered to accommodate the ionic radius. These chemical considerations complicate the picture, so it is better to first discuss the K-substituted phase diagram where this problem is not present, and then apply the conclusions to the Na version.

Potassium is slightly larger than barium (1.65 \r{A} vs. 1.56 \r{A}), and the KFe$_2$As$_2$ end member is a stable ThCr$_2$Si$_2$-type phase. There are two primary size considerations in substituting Ba with K: cation size and Fe oxidation state (or, put differently, the band filling of the (Fe$_2$As$_2$) layer). Replacing Ba$^{2+}$ with the slightly larger K$^+$ ion should expand the lattice, but it also affects the oxidation state of the Fe atom. The formal oxidation states are Fe$^2+$ in BaFe$_{2}$As$_{2}$ and Fe$^{2.5+}$ in KFe$_{2}$As$_{2}$. Thus, as Fe is oxidized by K substitution, the intralayer Fe-Fe spacing should shrink, which will decrease \emph{a} and also change the shape of the FeAs$_4$ tetrahedra. The Fe-As bond length stays almost constant as a function of composition (see below), which means that a decrease in Fe-Fe causes an increase in the width of the layer (illustrated schematically in Figure 6c). In general one would expect that the above electronic considerations will have a stronger effect on the \emph{ab} plane size than the A site cation, but that both will have an effect on the \emph{c} axis.  This is what is observed; as K is substituted for Ba, the \emph{c}-axis increases while the \emph{ab} area contracts. Indeed, the contraction in \emph{ab} area is also observed for other $A_{1-x}$Na$_x$Fe$_{2}$As$_{2}$ (\emph{A} = Sr, Ca, and Eu)\cite{19,20,21}, even though Na$^{1+}$ is larger than Ca$^{2+}$, which shows that the contraction is a direct result of the oxidation of the iron species. The net effect in Ba$_{1-x}$K$_x$Fe$_{2}$As$_{2}$ is actually a decrease in volume, though intriguingly at around 50\% substitution the values appear to become nearly constant. This may be due to the trade-off between local and itinerant states of the Fermi electrons, a balance that may change as the Fermi-level changes.

 \begin{figure*}
 \includegraphics[width=\textwidth]{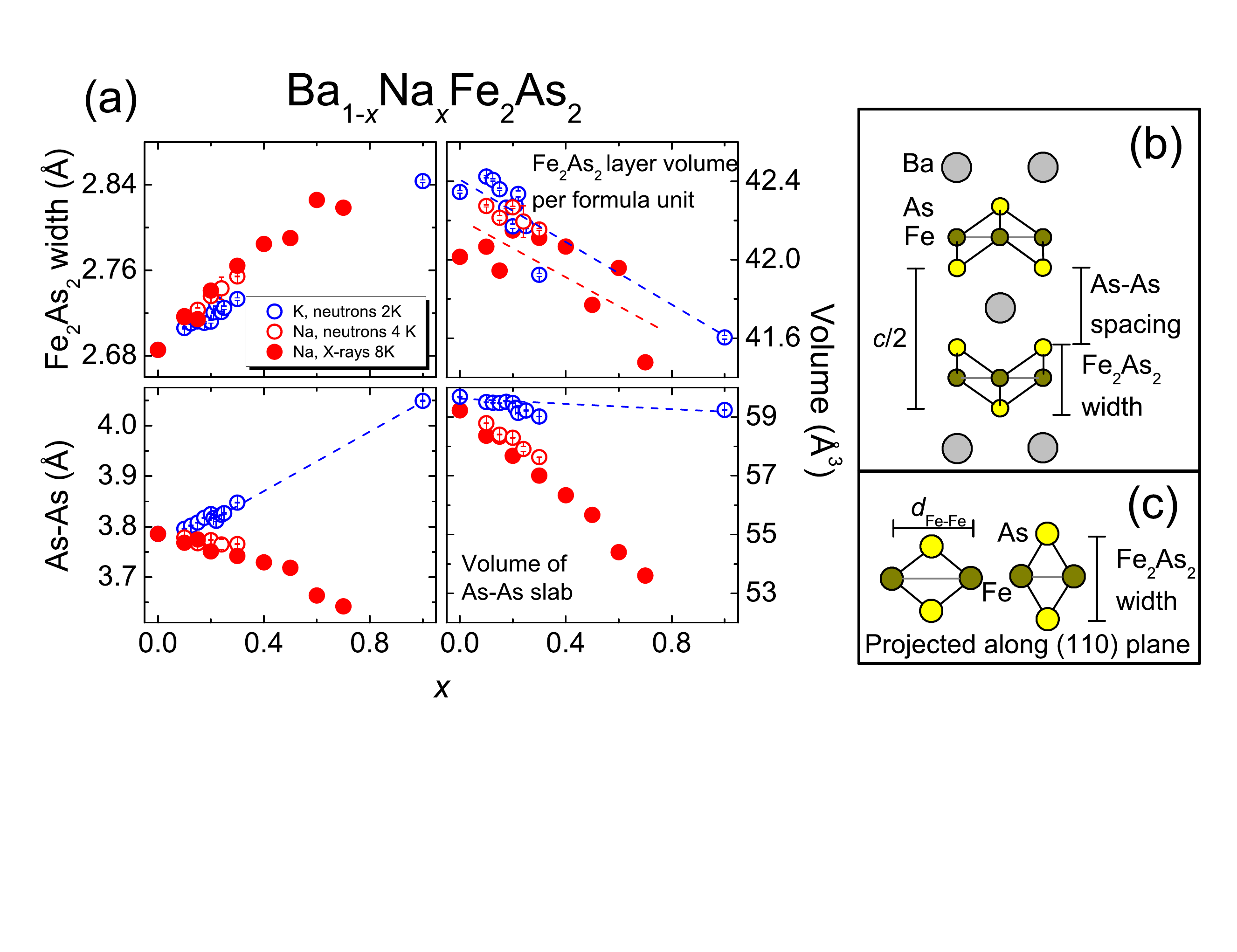}
\vspace{-115pt}
 \caption{(a) Layer widths (left panels) and volumes (right panels) for the Fe$_2$As$_2$ layer (top panels) and As-As spacing across the \emph{A} site (bottom panels). (b)  Depiction of how the \emph{c}-axis can be broken into As-As and Fe$_2$As$_2$ spacings. (c) Schematic for how Fe-Fe bond length affects the Fe$_2$As$_2$ width viewed along (110) direction.  The K-doped data is taken from Ref \cite{5}  \label{Figure6}}
 \end{figure*}

For the \emph{c}-axis itself, it can be understood as the sum of two physically meaningful parameters: the thickness of the Fe$_{2}$As$_{2}$ layer and the As-As separation between the layers (illustrated in figure 6b). The composition dependence of these parameters is plotted in figure 6a, left panels. The former is controlled by the bonding in the \emph{ab}-plane, and thus is most effected by atomic radius and bond lengths of the Fe and As species, while the latter is dominated by the As-As interaction and the \emph{A} cation size. Both of these increase approximately monotonically with K substitution, indicating that the As-As separation is sensitive to the K size (As-K bond length increases compared to As-Ba), while the increase in the Fe$_{2}$As$_{2}$ layer thickness indicates that oxidation of Fe broadens the tetrahedral layer. This is primarily steric: as the \emph{ab} plane shrinks, the As-Fe bond remains roughly the same length, so in order to keep the bond constant the As atoms must move out of plane (figure 6c). The layer width difference between BaFe$_{2}$As$_{2}$ and KFe$_{2}$As$_{2}$ is 0.158 \r{A} (5.9\%). For the As-As separation, it increases 0.263 \r{A} (7.0\%), which is quite substantial, yet the commensurate drop in cross-sectional area ($a^2$) yields a negligible volume change of 0.453 \r{A}$^3$ or 0.8\% (figure 6a, top right panel). Comparing the actual chemical separation around the \emph{A} site shows that Ba-As = 3.38 \r{A} and K-As = 3.40 \r{A}, which is somewhat smaller than the radial size differences between Ba and K. Taken together this implies that the overall interlayer spacing is mostly a steric effect, with decreases in a inducing increases in \emph{c} in order to keep an approximately constant \emph{A} site cage size. 

In the Na substituted system, the \emph{ab} plane shrinks a similar amount as it does in the K system for \emph{x} = 0.5, \emph{a} = 3.8894 and 3.8692 \r{A}, a decrease of 1.7 and 2.2\% for K and Na substitution, respectively. The larger decrease in the Na samples is expected as a consequence of the steric strain from the very small Na atom. For the \emph{c}-axis, the Fe$_{2}$As$_{2}$ layer thickness increases as it does in the K system, but the As-As separation shrinks slightly from $0 \leq x\leq 0.3$, and then starting around \emph{x} = 0.3 decreases dramatically. Plotting the volume of this slab shows a large, nearly linear decrease up to 5.63 \r{A}$^3$ (9.5\%) by \emph{x} = 0.7. Thus, the unusual shape of the As-As separation along the \emph{c} axis is merely an expression of the fact that the monotonically contracting a lattice parameter driven by the Fe$_{2}$As$_{2}$ layer has a non-linear effect on the \emph{A} site cage volume. Taking this together with the constantly increasing Fe$_{2}$As$_{2}$ layer width gives the overall \emph{c}-axis behavior that is observed. This means that despite their differences, the compositional lattice parameter trends in both systems can be understood as being primarily controlled by the Fe$_{2}$As$_{2}$ layer\textemdash itself controlled by the oxidation state of the iron\textemdash with secondary steric effects involving interlayer spacing compensating for the change in volume from the Fe-Fe shortening to accommodate the \emph{A} site cation size.

 \begin{figure}
\hspace{-25pt}
 \includegraphics[width=9cm]{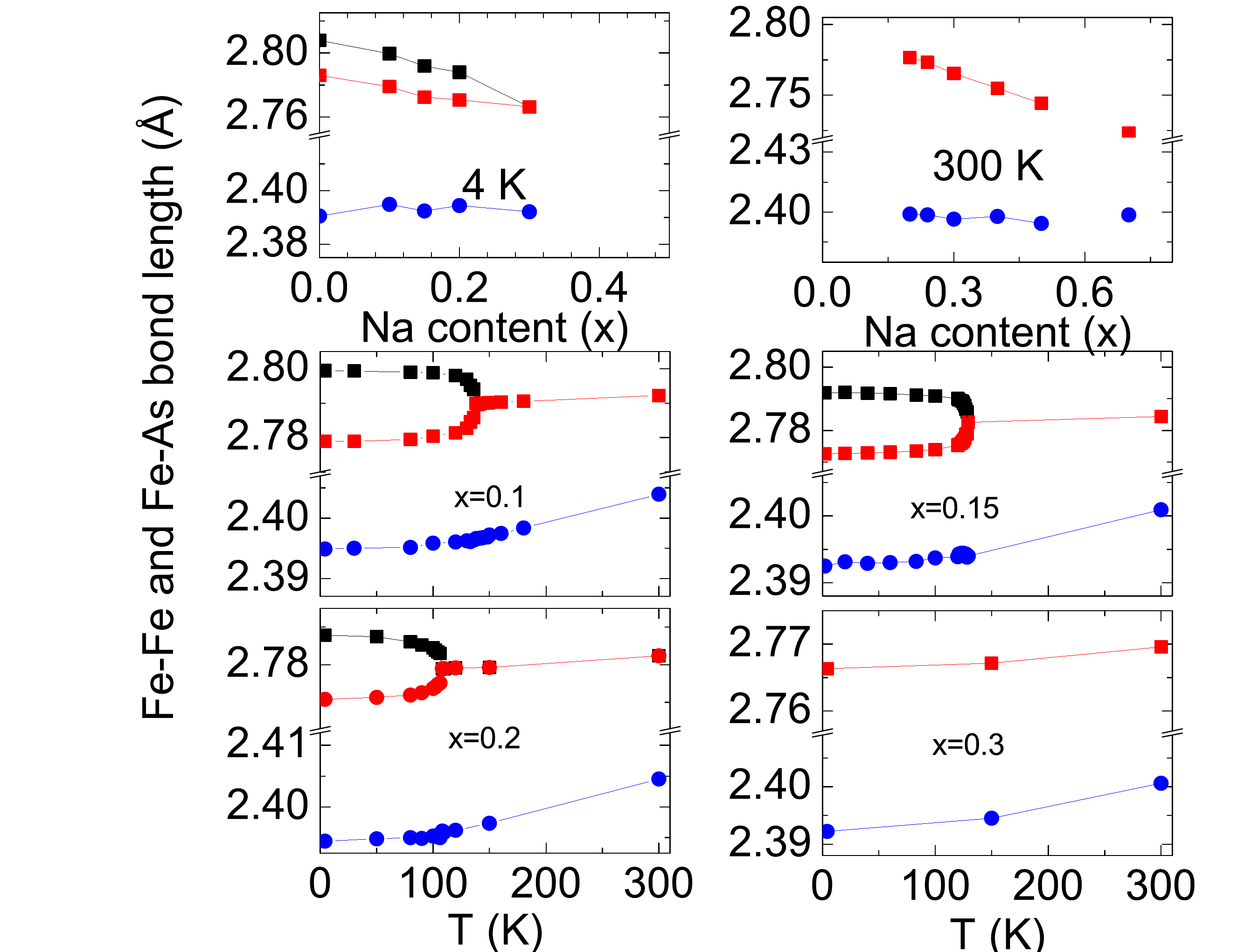}
 \caption{Variation of Fe-Fe and Fe-As bond lengths in Ba$_{1−x}$Na$_x$Fe$_2$As$_2$  with x at 4 K (neutron data) and 300 K (x-ray data), and with temperature (neutron data) for \emph{x} = 0.1, 0.15, 0.2, and 0.3. Blue circles represent the Fe-As bonds. Black and red squares represent the Fe-Fe bond lengths merging at $T_s$. Solid lines are guides for the eye. \label{Figure7}}
 \end{figure}

Na substitution and temperature dependence of the Fe-Fe and Fe-As interatomic distances are presented in Fig. 7. Behavior of the Fe-Fe distances is identical to that of the \emph{a} and \emph{b} lattice parameters because Fe atoms are located at invariant special positions in the unit cell. The Fe-As bond-length remains almost unchanged as a function of Na substitution both at room temperature and 4 K.  At a given composition it exhibits a very weak temperature dependence below the structural transition, but above $T_s$ the bond length varies much more strongly with temperature. In the K substituted system, we do not have data well above the structural transition temperature; however we previously reported that the Fe-As bond lengths remain unchanged up to the structural transition as in the Na substituted system.

 \begin{figure}
\hspace{-20pt}
 \includegraphics[width=9.2cm]{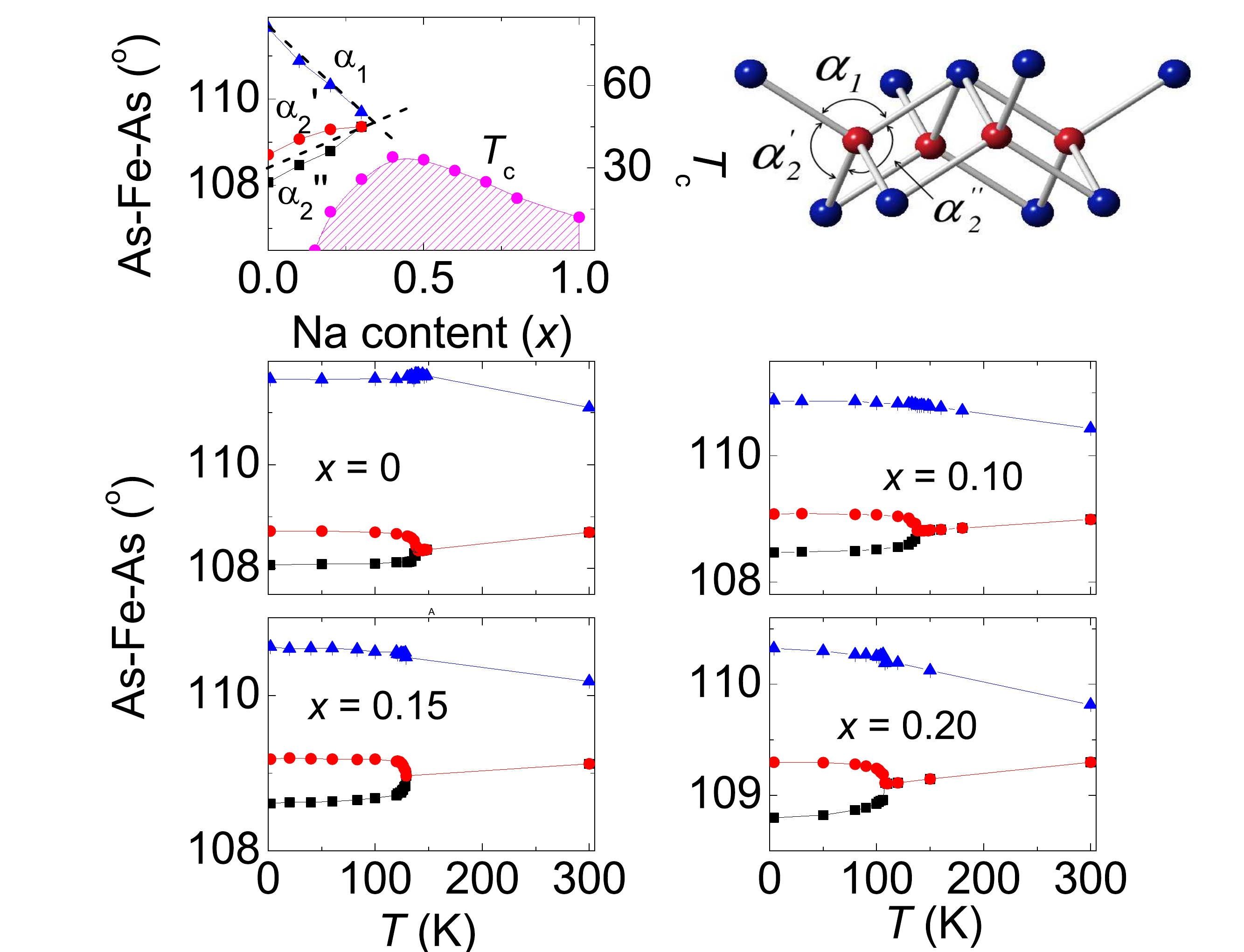}
 \caption{Variation of As-Fe-As bond angles in Ba$_{1−x}$Na$_x$Fe$_2$As$_2$  with \emph{x} at 1.7 K and with temperature. The top-right panel shows $\alpha_1$, $\alpha_2'$ and $\alpha_2''$ in orthorhombic setting. Blue triangles represent $\alpha_1$, red circles and black squares represent $\alpha_2'$ and $\alpha_2''$ merging into one $\alpha_2$ at $T_s$. Solid lines are guides to the eye. Comparison of angles and $T_c$ (pink shaded region) as a function of composition in shown in the top-left panel. Optimal $T_c$ occurs near perfect tetrahedral shape. \label{Figure8}}
 \end{figure}

Fig. 8 shows the As-Fe-As bond angle as a function of Na content and temperature. There are two independent angles, $\alpha_1$ and $\alpha_2$, in the tetragonal structure.  In the orthorhombic structure, the total number of independent angles increases to three, $\alpha_1$, $\alpha_2'$ and $\alpha_2''$ as shown in Fig. 8.  The top left panel includes the superconducting dome to show the relationship between the perfect tetrahedral angle and the maximum $T_c$. We reported a similar correlation in the K substituted system showing that the As-Fe-As bond angles are continuous when crossing from the electron-doped side to the hole-doped side of the phase diagram (see Fig. 12 of Ref \cite{5}) and the two independent angles $\alpha_1$ and $\alpha_2$ cross at $x \approx 0.4$ to yield a perfect tetrahedron and maximum $T_c$.  Kimber et al. \cite{12} studied the pressure dependence of the As-Fe-As bond angles in the BaFe$_{2}$As$_{2}$ parent compound and reported that the highest superconducting temperature of 31 K can be reached at 5.5 GPa, at which the As-Fe-As angles are very close to perfect tetrahedral angle. They proposed that the structural distortions that modify the Fermi surface play an important role in inducing superconductivity \cite{12}.

\subsection{Magnetism}
Fig. 9 shows the magnetization measurements for \emph{x} = 0, 0.1, 0.15 and 0.2 in a 2 kOe applied magnetic field. All four samples exhibit transitions associated with antiferromagnetic ordering. We have determined the N\'{e}el temperature values, $T_N$, as the maximum obtained from the first derivatives of the magnetization curves. These $T_N$ values were also confirmed by the neutron diffraction measurements.  

 \begin{figure}
 \includegraphics[width=8.6cm]{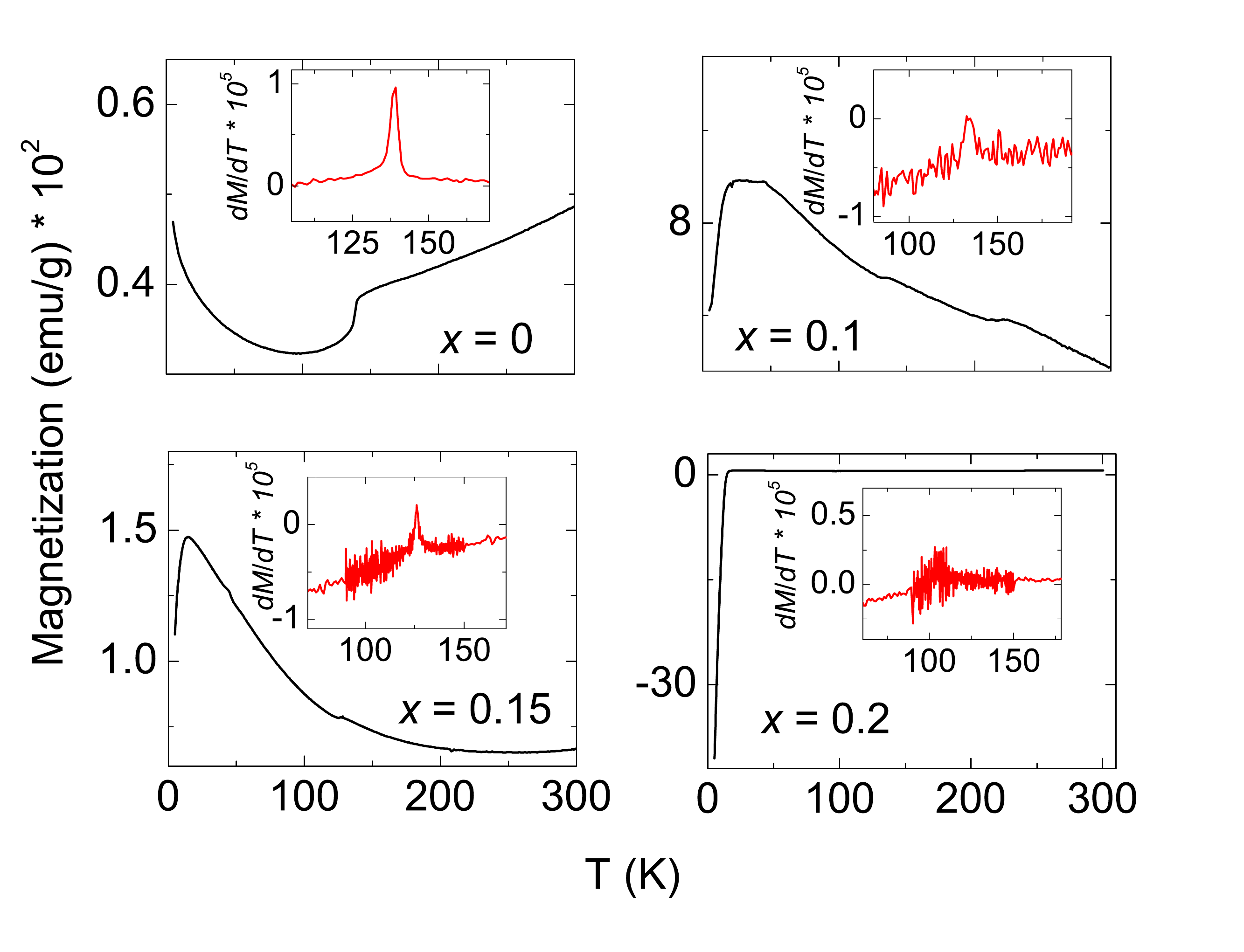}
\vspace{-23pt}
 \caption{SQUID magnetization measurements for \emph{x} = 0, 0.1, 0.15, and 0.2 in a 2-kG applied magnetic field. Insets are the first derivatives of the magnetization curves $dM/dT$ (10$^{-5}$) used to determine the N\'{e}el temperatures.\label{Figure9}}
 \end{figure}

Neutron powder diffraction data reveal the presence of tiny magnetic Bragg reflections below the N\'{e}el temperatures for all the orthorhombic $x < 0.3$ samples. Fig. 10 shows the neutron data above and below the magnetic transition for \emph{x}=0.1, 0.15, 0.2 and 0.3. The magnetic peaks (marked by arrows) located at 2.45 \r{A} and 3.43 \r{A} were indexed as (121) and (103) in agreement with the widely reported antiferromagnetic spin density wave (SDW) ground state.  \cite{8, 28,29}. These peaks are clearly present at base temperature for \emph{x} = 0.1, 0.15 and 0.2 samples, but for \emph{x} = 0.3 the intensity is nearly undetectable and the phase is tetragonal. With just this single diffractogram, it is difficult to state unambiguously whether there is magnetic order present. Nevertheless, this demonstrates the coexistence of superconductivity and SDW ground states in only a narrow compositional pocket of the phase diagram as was previously demonstrated for other 122 systems. For completeness, all possible magnetic symmetries associated with $Fmmm$ were tested but only the magnetic space group $F_Bm'm'm$ (the conventional setting is $F_Cmm'm'$) produced a good fit to the data. This is the same ground state that has been found in most of the other iron-pnictides: Fe magnetic moments couple antiferromagneticaly along the \emph{a}- and \emph{c}- axes and ferromagnetically along the \emph{b}-axis. \cite{29,30,31}

\begin{table}
\caption{Phase transition temperatures and 4 K magnetic and orthorhombic order parameters as determined by neutron powder diffraction. $T_s$ (the structural transition) and $T_{N,n}$ were extrapolated from a power law fit to the orthorhombic order parameters and refined magnetic moments, respectively. $T_{N,mag}$ is taken from the peak in $dM(T)/dT$ plot.\label{Table II}}
\begin{ruledtabular}
\begin{tabular}{c c c c c c}
\emph{x} & $T_s$ (K) & $T_{N,n}$ (K) & $T_{N,mag}$ (K) & \emph{M} ($\mu_B$) & $\delta*10^3$  \\
\hline
0.1 & 136.6(2) & 137.0(2) & 133.9(4) & 0.730(10) & 3.69(6) \\
0.15 & 128.1(3) & 128.1(3) & 126.1(1) & 0.728(13) & 3.45(2) \\
0.2 & 107.0(1) & 107.0(1) & 104.6(7) & 0.669(10) & 3.05(5) \\
\end{tabular}
\end{ruledtabular}
\end{table}

 \begin{figure}[b]
\vspace{-15pt}
\hspace{-0.8cm}
 \includegraphics[width=9.4cm]{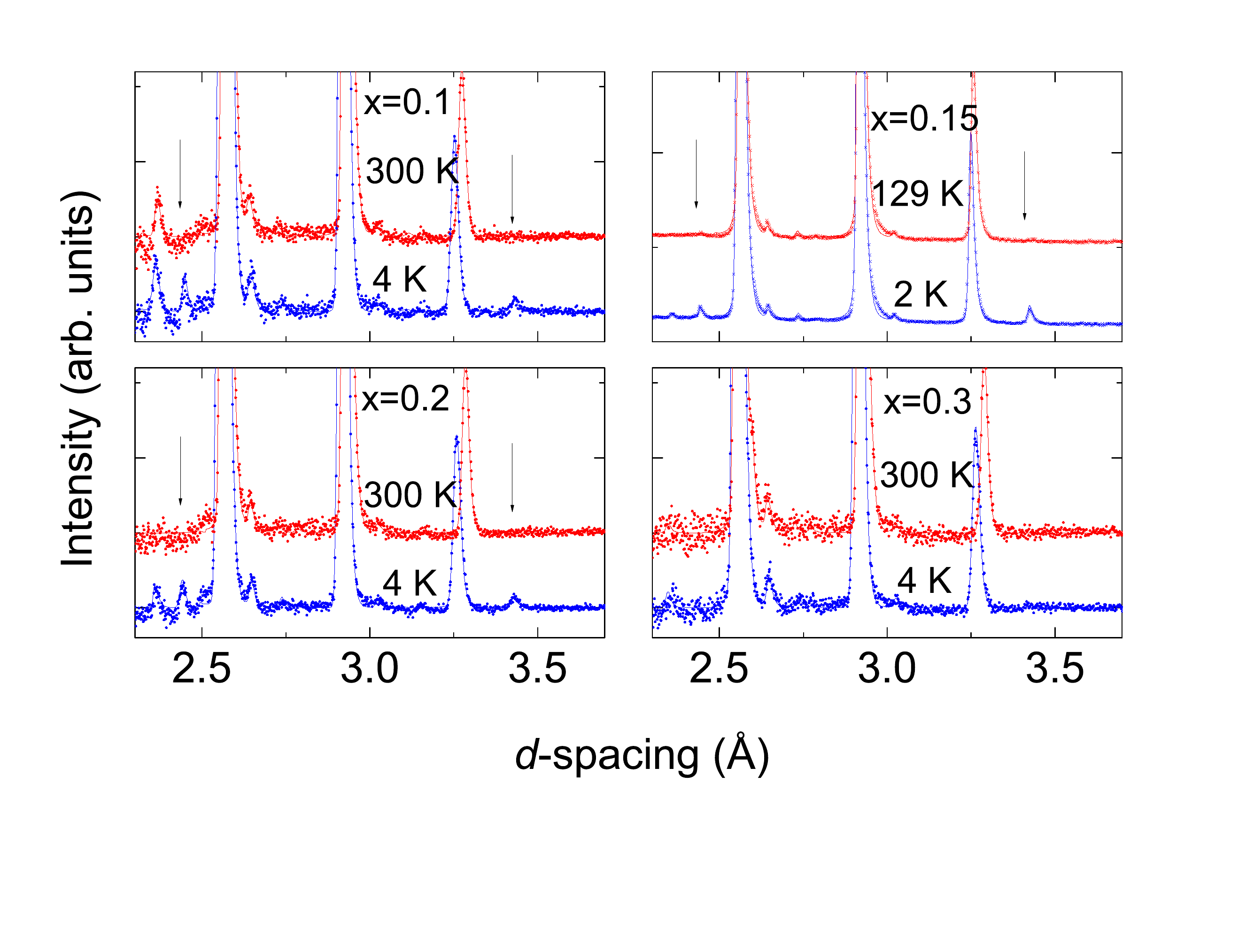}
\vspace{-50pt}
 \caption{Neutron diffraction from Ba$_{1-x}$Na$_x$Fe$_2$As$_2$  at 1.7 K with the magnetic Bragg peaks at a \emph{d}-spacing of 2.45 and 3.43 \r{A} indicated by the arrows. They are absent above $T_N$ for \emph{x} = 0.1, 0.15, and 0.2. At \emph{x} = 0.3, a nearly negligible amount of intensity is detected above background. \label{Figure10}}
 \end{figure}

Fig. 11 shows the orthorhombic and magnetic order parameters as a function of temperature for \emph{x} = 0, 0.1, 0.15 and 0.2. The identical behavior of these two independently determined order parameters demonstrate the existence of a strong magneto-elastic coupling that forces the magnetic and structural transitions to occur simultaneously.  It is important to note that the orthorhombic and magnetic order parameters were independently determined from the same neutron diffraction data by the splitting of the nuclear Bragg peaks and by the intensity of the magnetic Bragg peaks, respectively. This excludes any spurious issues related to the accuracy of the thermometry used in the experiment.  We reported an identical behavior in the K substituted system, which supports the idea that the magnetically driven structural transition is a generic feature of the hole-doped regime. The magnetic moment, $T_N$ as determined from the power-law fits to magnetic moment, $T_N$ as determined from magnetization, and $T_s$ as determined from power-law fits to the orthorhombic order parameters are all tabulated in Table II for comparison.

 \begin{figure}
 \includegraphics[width=8.6cm]{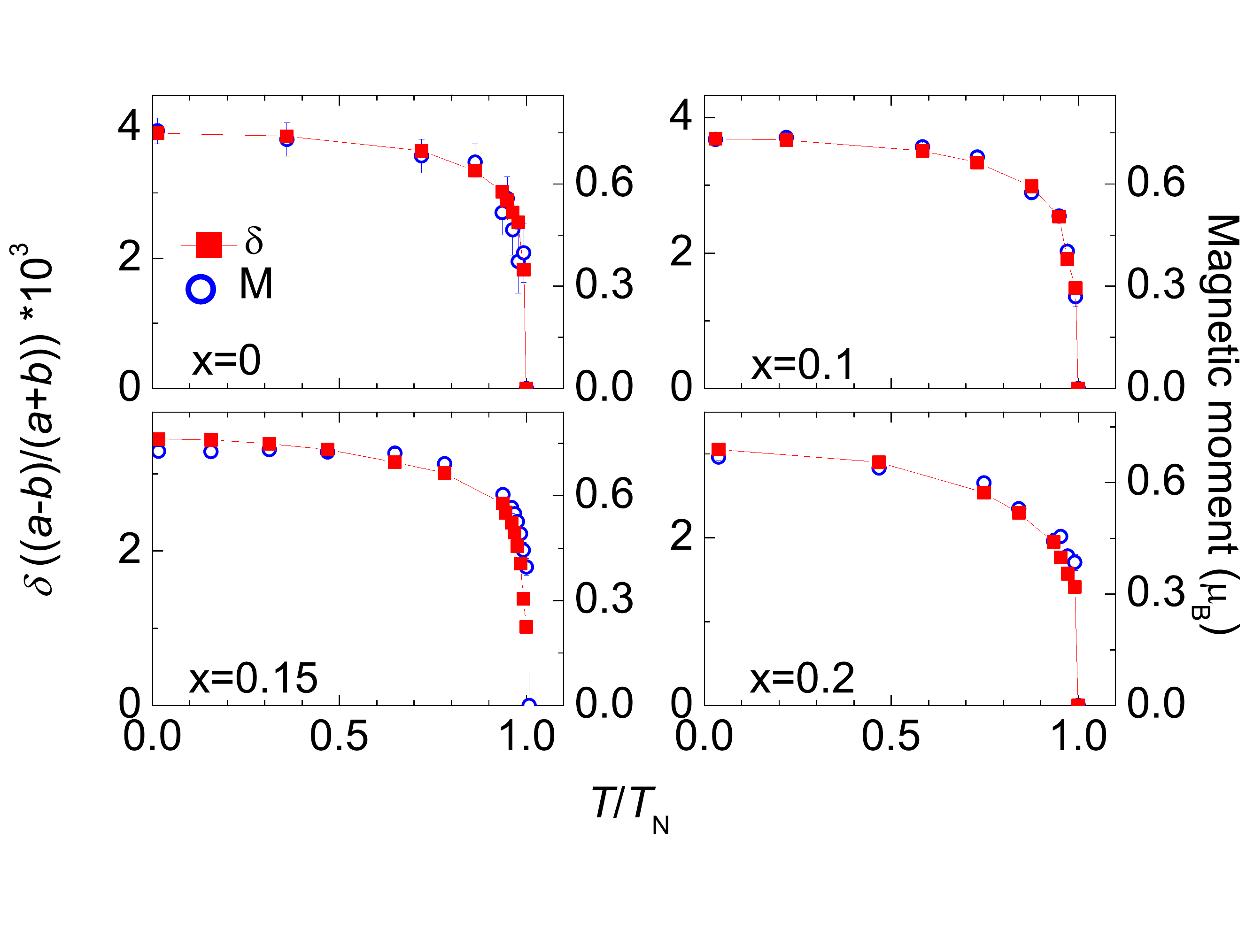}
\vspace{-40pt}
 \caption{Temperature dependences of the orthorhombic (solid red squares) and magnetic (open blue circles) order parameters for \emph{x} = 0, 0.1, 0.15, and 0.2 samples. \label{Figure11}}
 \end{figure}

\section{Conclusion}
 A detailed phase diagram for Ba$_{1-x}$Na$_x$Fe$_2$As$_2$ is displayed in Fig. 12 in which our high quality samples allow for the proper determination of continuous and smooth phase boundary lines that separate the paramagnetic tetragonal phase from the superconducting dome and the antiferromagnetic SDW regime. Unlike the Co-substituted Ba-122 system in which the magnetic and structural transitions are separated, the transitions in Ba$_{1-x}$Na$_x$Fe$_2$As$_2$ are coincident and first-order as was also observed for the K-substituted Ba-122.  A narrow pocket of overlapping superconductivity and antiferromagnetism is present within the composition range $0.15< x <0.3$. In the K-substituted system we reported ~5 \% reduction in the orthorhombic and magnetic order parameters below the superconducting transition temperature for the samples that show both antiferromagnetic ordering and at least 80 \% superconducting volume fraction \cite{4}. From this result, we concluded that in the K-substituted system, superconductivity and magnetism are microscopically coexistent, otherwise the magnetic order parameter would have been suppressed at least 80 \% below $T_c$. In the \emph{x} = 0.24 sample of the Na-substituted system, we observed a similar behavior that the integrated intensity of the (101) magnetic peak is suppressed by ~25 \% below 18 K. This composition is included in the phase diagram as an AF/T regime \cite{22}. The phase boundaries are currently unknown, so it is depicted here conservatively as a very narrow dome peaking near $x = 0.24$. That sample is reported as biphasic: it contains both AF/T and AF/O, and both phases appear to be superconductors.

\begin{figure}
 \includegraphics[width=8.6cm]{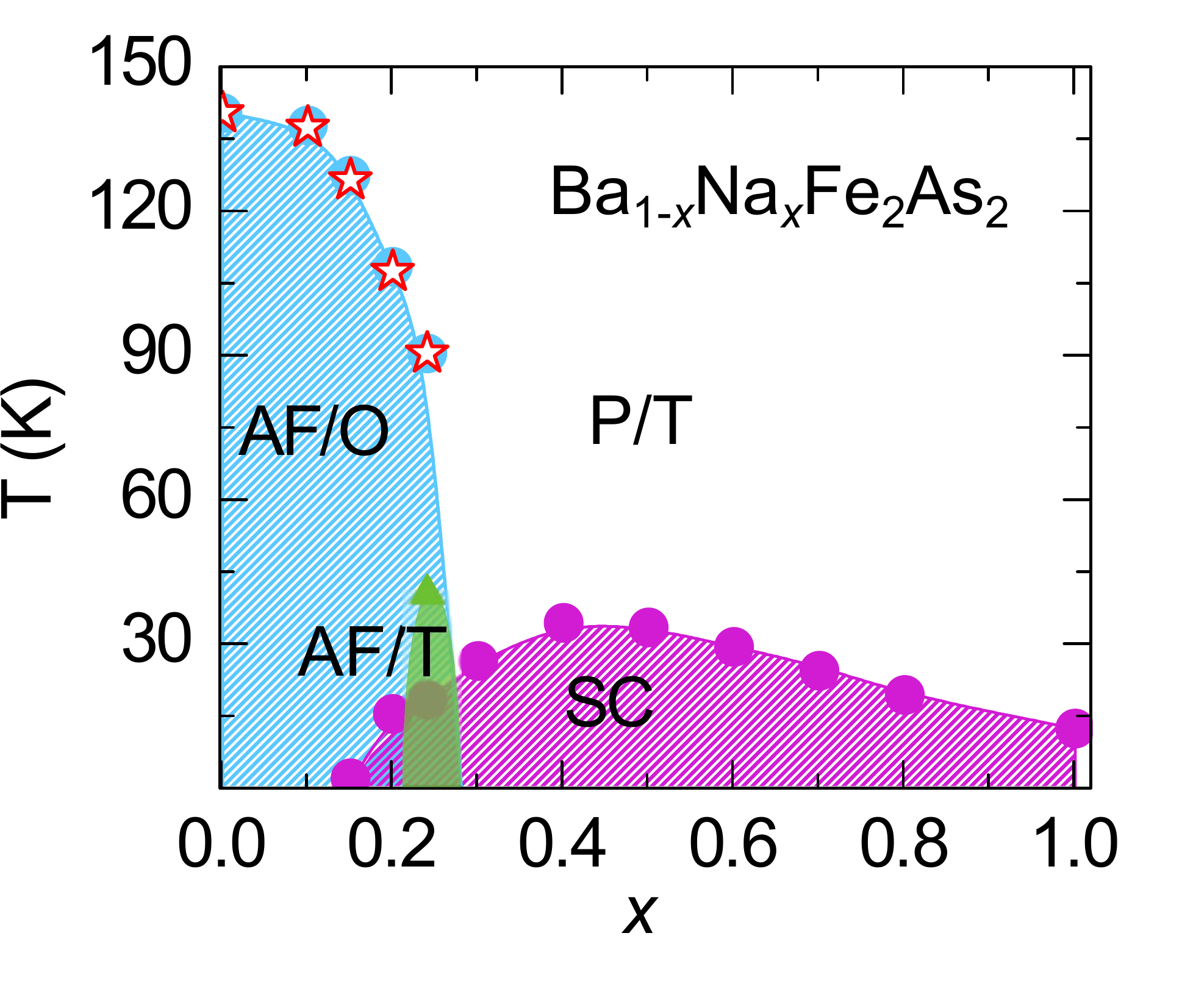}
\vspace{-25pt}
 \caption{Phase diagram of Ba$_{1-x}$Na$_x$Fe$_2$As$_2$. P/T is the normal state paramagnetic tetragonal phase and AF/T refers to the antiferromagnetic tetragonal phase (C4 in the text) [25]. The \emph{x} = 1.0 point is taken from \cite{22} \label{Figure12}}
 \end{figure}

The fact that the phase diagram of the Na-doped materials bears such a strong resemblance to the K-doped phase diagram despite the very different ionic sizes has some important implications for the relationship between nuclear structure, electronic structure, and properties. Despite the different behaviors observed in the \emph{c} axis for each series, the \emph{a},\emph{b} behavior is remarkably similar. This suggests that for these hole doped systems the Fe$_2$As$_2$ layer shape is only sensitive to the \emph{A} site cation size as a secondary effect, while the primary consideration is the Fe$_2$As$_2$ layer’s overall composition including electron count. This also agrees with the fact that pressure-induced superconductivity in underdoped materials has been found to be maximized by strain in the \emph{ab} plane, while the \emph{c}-axis strain inhibits superconductivity \cite{32}.

The materials’ properties primarily depend on the Fe$_2$As$_2$ shape, which is why both electronic phase diagrams express so many similarities, but it also explains why the small changes in the layer caused by the chemical pressure from the Na vs. K doping can change the shape of the superconducting dome slightly and introduce the unexpected property of a reentrant magnetic phase in the Na system under ambient pressure.

\section{Acknowledgments}
Work at the Materials Science Division and Advanced Photon Source at Argonne National Laboratory was supported by the U. S. Department of Energy, Office of Science, Office of Basic Energy Sciences, under Contract No. DE-AC02-06CH11357. Work at the Lujan Center at Los Alamos Neutron Science Center, was funded by DOE Office of Basic Energy Sciences under DOE Contract DE-AC52-06NA25396. Experiments at the ISIS Pulsed Neutron and Muon Source were supported by a beam time allocation from the Science and Technology Facilities Council.

%
\end{document}